\documentclass[showpacs,amssymb,amsmath,nobibnotes, aps,prl,secnumarabic
,pdfoutput %,superscriptaddress
,twocolumn%
]{revtex4-1}

\usepackage{bm}
\usepackage{natbib}
\usepackage{graphicx}

\begin{document}
\textfloatsep 10pt

\title{Taylor particle dispersion during transition to fully developed two-dimensional turbulence}
\author{H. Xia}
\email{Hua.Xia@anu.edu.au}
\author{N. Francois}
\author{H. Punzmann}
\author{M. Shats}

\affiliation{Research School of Physics and Engineering, The Australian National University, Canberra ACT 0200, Australia}

\date{\today}

\begin{abstract}
We report new measurements of single particle dispersion in turbulent two-dimensional (2D) flows. Laboratory experiments in electromagnetically driven and Faraday wave driven turbulence reveal a transition from weakly dispersing superdiffusive regime to strongly dispersing Brownian diffusion as the flow energy is increased in a broad range. The transition to fully developed 2D turbulence is characterized by the topological changes in the fluid particle trajectories and the development of self-similar diffusion. The degree of 2D turbulence development can be quantified by a parameter describing the deviation of single particle dispersion from the Taylor dispersion.
\end{abstract}


\maketitle
Quantifying particle dispersion in disordered and turbulent flows is a great challenge in oceanographic and  atmospheric  applications \cite{LaCasce2008, Dosio2005}, in the development of the sea search and rescue algorithms \cite{Breivik2013}, and in industrial mixing \cite{Mixing2004}. The theory of Lagrangian statistics of particle displacements by Taylor (1921) \cite{Taylor1921} gives a mean squared displacement (MSD) $\langle \delta r^2 \rangle =\langle [ \vec{r}(t)-\vec{r}(0) ]^2  \rangle$ of a particle moving along the trajectory $\vec{r}(t)$ from its initial position $\vec{r}(0)$ in a turbulent flow as:
\begin{equation}\label{Taylor}
    \begin{array}{cc}
       \langle \delta r^2 \rangle = \tilde{u}^2 t^2,  & t < T_L \\
       \langle \delta r^2 \rangle = 2D t,  & t > T_L
     \end{array}
\end{equation}
\noindent where $\tilde{u}^2$ is the velocity variance, and $T_L=\int_0^\infty \rho(t) dt$ is the Lagrangian integral time which can be obtained by integrating the Lagrangian velocity autocorrelation function
\begin{equation}\label{ACF}
\rho(t)=\langle \textbf{u}(t_0+t) \textbf{u}(t_0) \rangle / \tilde{u}^2.
\end{equation}
The first equation (Eq.~\ref{Taylor}) describes the ballistic motion of particles at short times, while the second one gives the Brownian-type diffusion where $D=\tilde{u}^2 T_L$.

In many geophysical and laboratory flows diffusion is anomalous, i.e. at long times the MSD is not a linear function of time, such that $\langle \delta r^2 \rangle \sim t^{\gamma_2}$, where $\gamma_2 \neq 1$ ($1<\gamma_2 < 2$ is called superdiffusion, while $\gamma_2 < 1$ corresponds to a subdiffusive process). Superdiffusion is a non-Brownian diffusion process in which particles experience a succession of small and very large displacements resulting in the tails of the probability density functions of the particle displacement, and a slow decay of the autocorrelation function of Lagrangian velocities \cite{Castiglione1999, Ferrari2001}. Anomalous diffusion regimes have been successfully described using generalized models, such as L\'{e}vy walks \cite{Shlesinger1986,Shlesinger1987} or truncated L\'{e}vy flights \cite{Mantegna1994}. Brownian walk is a special member of the L\'{e}vy flight random walk \cite{Klafter1996}. L\'{e}vy flights have been studied in chaotic flows in rotating fluids \cite{Solomon1993, Weeks1996}. In developed 2D turbulence single particle dispersion quantitatively agrees with Taylor's result (Eq.~\ref{Taylor}) as was recently demonstrated in experiments \cite{XiaNC2013}. It was found that the diffusion coefficient in turbulence can be expressed as $D=\tilde{u}L_L$, Here $\tilde{u}$ is the r.m.s. of the velocity fluctuations and the Lagrangian correlation length $L_L$ is related to the turbulence forcing scale $L_f$ as $L_L \approx (\sqrt{2}/2) L_f$. At lower flow energies however deviations from this relation are noticeable.

In this Letter the particle dispersion is investigated for the first time in a broad range of the Reynolds numbers, or kinetic energies. The motivations for this are as follows. It is sometimes argued that 2D turbulence modeled in laboratory cannot be relevant in the analysis of geophysical flows since the Reynolds numbers achieved in laboratory are by orders of magnitude smaller than in geophysics. Moreover, there is no well accepted parameter to quantify the degree of turbulence development in 2D and it is not clear whether the Reynolds number is a meaningful measure of it. Indeed, in 2D turbulence energy is transferred from the forcing scale towards larger scales, where the role of viscous dissipation is low, such that the meaning of the Reynolds number is very different from 3D turbulence.

Here we propose to characterize the degree of turbulence development from the Lagrangian perspective.
To quantitatively characterize deviations of the single particle dispersion from a random walk represented by Eq.~(\ref{Taylor}), it is not sufficient to establish the anomalous scaling $\langle \delta r^2 \rangle \sim t^{\gamma_2}$. In addition to $\gamma_2$, one needs to compare the values of the MSD with Taylor's prediction. Here we show that in a broad range of the flow kinetic energies, the MSD can be expressed as
\begin{equation}\label{MSD}
\frac{\langle (\delta r)^2 \rangle}{L_f^2}=\beta\left(\frac{t}{T_L}\right)^{\gamma_2},    t>T_L
\end{equation}
\noindent where $\beta$ is a new measure of the particle dispersion anomaly. It is strongly dependent on the flow kinetic energy $E=(1/2) \tilde{u}^2 $ and it is much more sensitive to the degree of the turbulence development than $\gamma_2$. The results reveal a clear transition from superdiffusion in flows at lower $E$ characterized by $\beta <1$ and $\gamma_2>1$, to a Brownian diffusion regime in the developed 2D turbulence with $\beta=\gamma_2=1$. The new empirical law allows a quantitative estimation of the diffusion coefficient. The results point to the existence of a clearly detectable criterion for the transition to Brownian-type diffusion in 2D turbulence. We propose that the value of the coefficient $\beta$ in Eq.~(\ref{MSD}) is a sensitive new measure of the 2D turbulence development.

In the experiments turbulence is produced using two distinctly different methods. Electromagnetically driven turbulence (EMT) is produced in layers of electrolyte by running electric current across the fluid cell (square container $30 \times 30$ cm) placed above the array of permanent magnets \cite{Xia2008,Xia2009,XiaNP2011,XiaNC2013}. A double layer configuration is employed to reduced the bottom dissipation and 3D effects~\cite{Shats2010}, where 4mm thick $Na_2SO_4$ water solution is placed on top of 4mm of heavier, non-conducting fluid (FC-3283). The Lorenz force produces vortices which interact with each other, thus generating complex quasi-2D flow. By changing the current density one can control the degree of turbulence development and the energy injected into the flow at the scale which is approximately equal to the distance between the magnets (9 mm in this experiment), the forcing scale $L_f$.

The second method of turbulence generation, the Faraday wave turbulence (FWT), was discovered recently on the surface of the vertically vibrated liquids \cite{vonKameke2011,Francois2013}. The motion of particles on the surface perturbed by parametrically excited Faraday waves remarkably reproduces fluid motion in 2D turbulence. In these experiments Faraday waves are generated in a 180 mm diameter circular container filled with water which is periodically shaken at the frequency of 60 Hz at  the peak-to-peak acceleration in the range $a=(0.7-2)g$. The lowest acceleration is determined by the threshold of parametric wave excitation, while the highest is limited by the droplet formation.

The two different schemes of turbulence generation are used to ensure the findings on the particle dispersion in this paper are independent of the specific method of flow generation.In both experiments, the motion of the fluid surface is visualized by placing 50 $\mu$m polyamid particles on the water surface. The use of surfactant and plasma treatment of the particles ensures homogenous distribution of the tracer particles. The particle motion is captured using high-resolution fast camera (Andor Neo sCMOS) as described in \cite{Francois2013,XiaNC2013}.

Table \ref{tab:par} summarizes the experimental conditions studied here. Turbulence forcing (2nd column) is controlled by changing the electric current density in the EMT and by changing the vertical acceleration in the FWT. The increase in forcing leads to the increase in the horizontal kinetic energy (horizontal velocity variance $\tilde{u}^2=2E$ is shown in the 3rd column). The forcing scale Reynolds number defined as $Re = \tilde{u} L_f/\nu$ (where $\nu$ is the kinematic viscosity) is shown in the last column of table \ref{tab:par}. The ranges of the $Re$ and of the $\tilde{u}^2$ overlap substantially in the two experimental schemes. The highest achievable $Re$ is restricted in the EMT by the maximum electric current density (ohmic heat dissipation in the electrolyte), while in the FWT it is limited by the droplet generation threshold. The range of the flow energies is  so far the largest studied in laboratory experiments; this has been made possible due to the development of a new method of 2D turbulence generation using Faraday waves \cite{vonKameke2011,Francois2013}.

\begin{table}
  \begin{center}
\def~{\hphantom{0}}
  \begin{tabular}{lcccc}
      Label  & Forcing   & $\tilde{u}^2$ ($m^2/s^2$) & $L_f$ ($mm$) &   $Re$\\[3pt]
      \hline
  EMT1 & $0.1\times10^3$ A/m$^2$ &        $1.48 \times 10^{-6}$ &9& 11 \\
  EMT2 & $0.2\times10^3$ A/m$^2$ &$4.56 \times 10^{-6}$&9&  19 \\
  EMT3 & $0.4\times10^3$ A/m$^2$ &$7.6 \times 10^{-6}$&9&  25 \\
  EMT4 & $0.5\times10^3$ A/m$^2$ &$1.6 \times 10^{-5}$&9&  36 \\

  EMT5 & $0.6\times10^3$ A/m$^2$ &$2 \times 10^{-5}$&9&  40 \\
  EMT6 & $0.8\times10^3$ A/m$^2$ &$3.1 \times 10^{-5}$&9&  50 \\
  EMT7 & $1.\times10^3$ A/m$^2$ &$3.6 \times 10^{-5}$&9&  56 \\

  EMT8 & $1.2\times10^3$ A/m$^2$ &$5.1 \times 10^{-5}$&9&  64 \\
  EMT9 & $1.5\times10^3$ A/m$^2$ &$8.8 \times 10^{-5}$&9&  80 \\

  FWT1 & 60Hz, 0.7g &$1.24 \times 10^{-5}$&4.4&  15 \\
  FWT2 & 60Hz, 0.8g &$3.24 \times 10^{-5}$&4.4&  25 \\
  FWT3 & 60Hz, 0.9g &$8.9 \times 10^{-5}$&4.4&  42 \\
  FWT4 & 60Hz, 1.0g &$1.6 \times 10^{-4}$&4.4&  56 \\
  FWT5 & 60Hz, 1.2g &$3.2 \times 10^{-4}$&4.4&  79 \\
  FWT6 & 60Hz, 1.4g &$6.1 \times 10^{-4}$&4.4&  108 \\
  FWT7 & 60Hz, 1.6g &$1.02 \times 10^{-3}$&4.4& 140 \\
  FWT8 & 60Hz, 1.8g &$1.46 \times 10^{-3}$&4.4& 168 \\
  FWT9 & 60Hz, 2.0g &$1.83 \times 10^{-3}$&4.4& 188 \\

  \end{tabular}
  \caption{Experimental parameters for different experiments analyzed here. }
  \label{tab:par}
  \end{center}
\end{table}

\begin{figure}
  \centerline{\includegraphics[width=8.5cm]{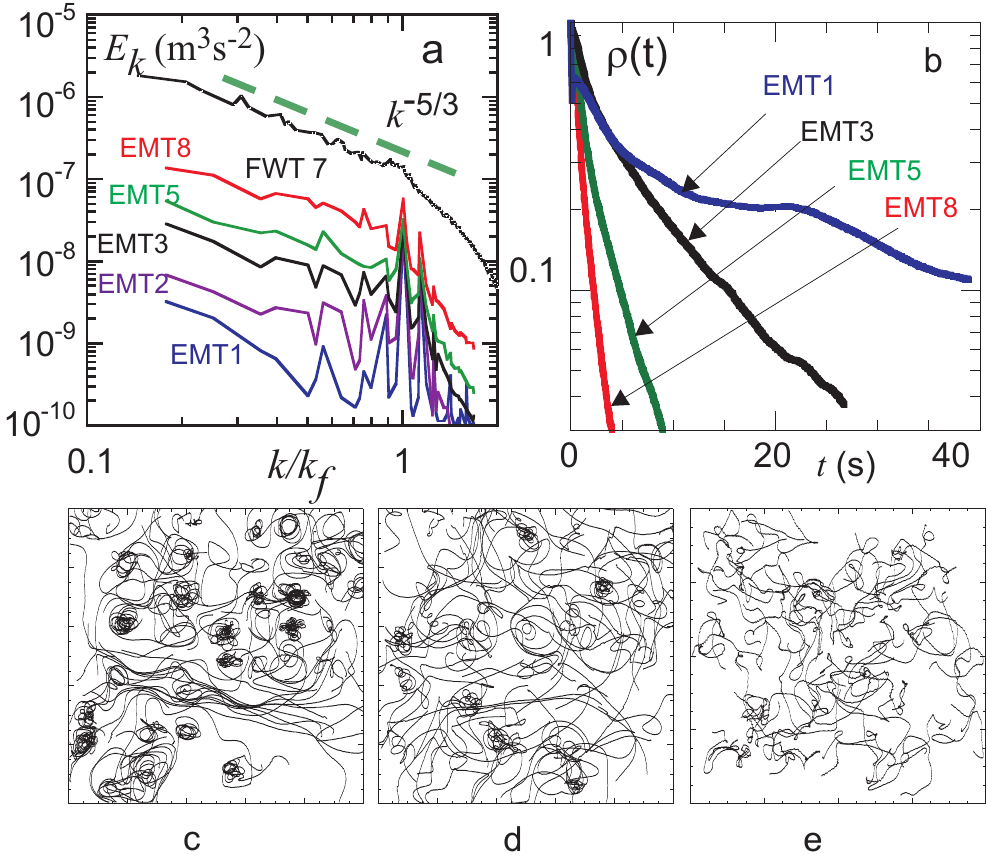}}
  \caption{(a) Eulerian wave number spectra of kinetic energy $E(k)$ in the EMT and in the FWT measured at different forcing levels (see Table 1 for labeling). (b) Lagrangian velocity autocorrelation function $\rho(t)$ for different experiments. Examples of fluid particle trajectories in (c) EMT3 ($Re=25$), (d) EMT8 ($Re=64$), (e) FWT5 ($Re=79$). The box size is 10cm and 8cm for EMT and FWT, respectively.}
\label{fig1}
\end{figure}

The trajectories of particles are tracked using a nearest neighbor algorithm \cite{Crocker1996}. Each 2D particle track gives the Lagrangian velocities along the trajectory. The measurement is performed twice with two different seeding densities of particles: lower for the particle tracking, and higher for the particle image velocimetry. The horizontal kinetic energy of the flow is measured using both the Lagrangian and Eulerian velocities, which shows the same results. The latter is used for generating 2D velocity fields on a grid (typically 90$\times$90, which can also be interpolated). Figure~\ref{fig1}(a) shows the kinetic energy spectra $E_k$ versus the normalized wave number $k/k_f$ for several experiments. It shows that the increase in the Reynolds number leads to the development of the inverse energy cascade, i.e. the energy flows towards lower wave numbers, $k<k_f$. The  spectrum gradually converges at higher $Re$ to  a well known Kolmogorov-Kraichnan spectrum  $E_k=C\epsilon^{2/3} k^{-5/3}$ . Here $C \approx 6$ and $\epsilon$ is the energy dissipation rate (see \cite{XiaNC2013}). Though the development of the broad $k^{-5/3}$ spectrum is indicative of the degree of turbulence development, we show below that the Lagrangian statistics are more sensitive and can be used to define the transition to fully developed 2D turbulence.

The autocorrelation functions of the Lagrangian velocity $\rho(t)$ (measured along the trajectories) change with the increase in forcing as seen in Fig.~\ref{fig1}(b). At the lowest forcing they exhibit strong non-exponential tails, which gradually disappear, and the autocorrelation functions converge to decaying exponentials $\rho(t)=$exp$(-t/T_L)$. The particle trajectories shown in Fig.~\ref{fig1}(c-e) illustrate the change in the nature of the particle motion at different forcing. At low forcing, Fig.~\ref{fig1}(c,d), trajectories somewhat resemble L\'{e}vy flights, seen as small displacements of particles trapped within the forcing scale vortices for a long time, followed by large jumps. Trapped particles are responsible for long tails in the autocorrelation functions, while the jumps, or flights, represent memory loss events as particles move from one vortex to another. Similar "sticking and flight" behavior has been observed in chaotic rotating fluid experiments \cite{Solomon1993} and also in the particle motion driven by Faraday waves at relatively low drive \cite{Hansen1997}. As the turbulence energy increases, Fig.~\ref{fig1}(e), time spent by fluid particles within the traps is reduced and trajectories no longer show forcing scale vortices.

The particle displacements $\delta r$ and their moments $\langle|\delta r|^p\rangle$ are computed over several thousands of trajectories for each experiment. The degree of the diffusion self-similarity can be determined from the higher order moments of the particle displacements \cite{Ferrari2001}. If $\langle |\delta r|^p\rangle \sim t^{\gamma_p}$ for $t>T_L$, where $\gamma_p = p/2$, the particle dispersion is said to be strongly self-similar and diffusive. Otherwise, if $\gamma_p \neq p/2$ and $\gamma_p$ scales as a nonlinear function of $p$, for example exhibiting a piece-wise linear dependence, the diffusion is said to be non-self-similar. Exceptional displacements, and tails in the probability density function (PDF) of $\delta r$ are expected from theoretical models in non-self-similar diffusion \cite{Castiglione1999, Ferrari2001}. However, neither the higher order moments of the particle displacements, nor PDFs had been investigated before in laboratory turbulence.

\begin{figure}
  \centerline{\includegraphics[width=8.5cm]{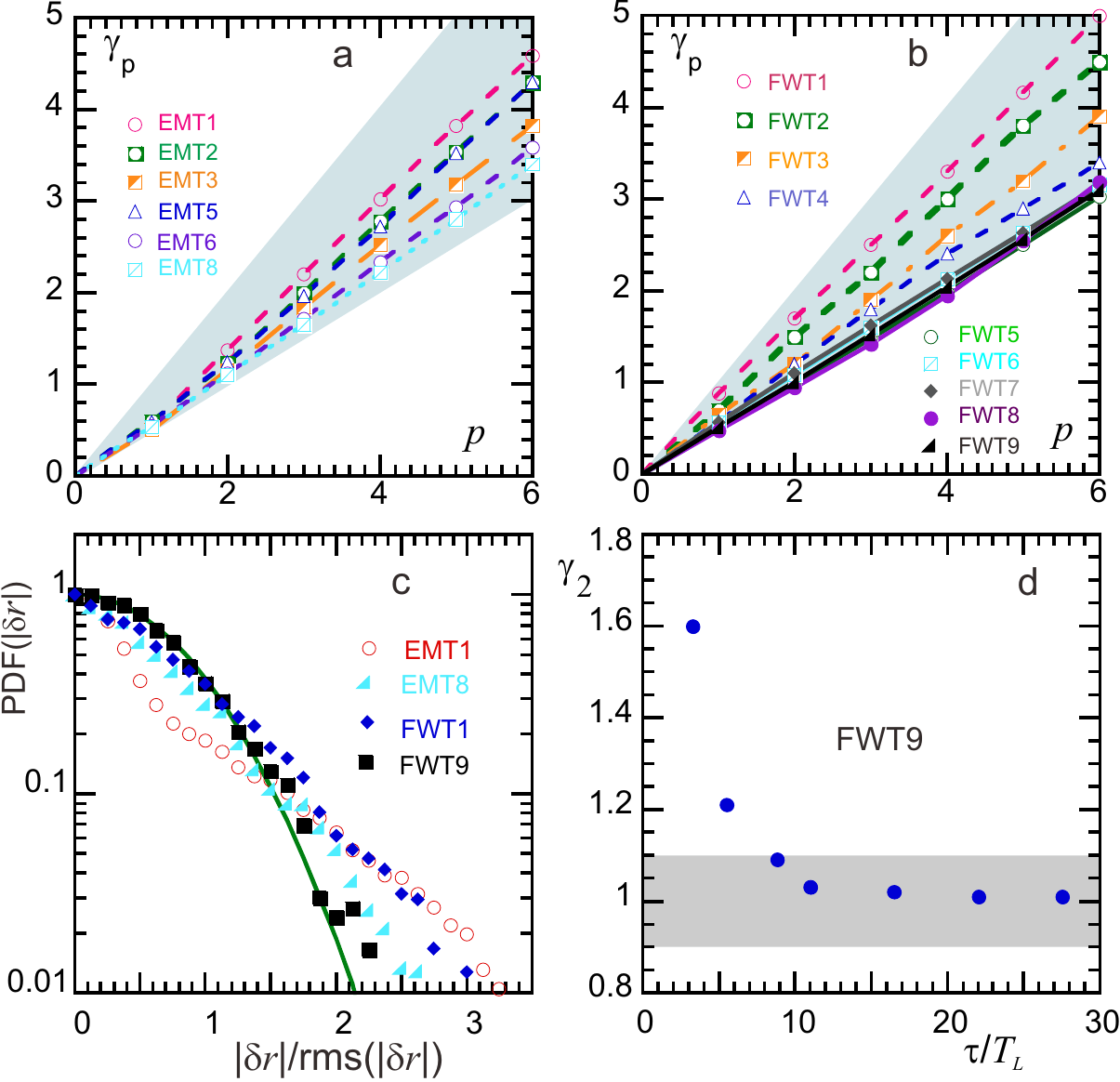}}
  \caption{$\gamma_p$ as a function of $p$ in (a) EMT1-8, and (b) FWT1-9. The shaded area shows the range between self-similar diffusion $\gamma_p = p/2$ and ballistic particle displacement $\gamma_p = p$. (c) Probability density functions of particle displacements versus $\delta r$ normalized by their r.m.s. value in EMT1 (open circles), EMT8 (grey triangles), FWT1 (solid diamonds) and FWT9 (solid squares). Solid line shows a Gaussian fit. (d) Second-order moment $\gamma_2$ computed using trajectories of different length. The length of the trajectory is expressed as tracking time $\tau$ normalized by the Lagrangian integral time $T_L$.  }
\label{fig2}
\end{figure}

Figures~\ref{fig2}(a,b) show the scaling $\gamma_p$ of the first 6 moments of the particle displacement $|\delta r|$ as a function of $p$ for different EMT and FWT flows. The highest Reynolds number flows (FWT) show self-similar diffusion with $\gamma_p = p/2$. However the lower $Re$ flows show substantial deviations from this trend, especially for the higher moments.

Correspondingly, the particle displacement PDFs exhibit different functional forms. Figure~\ref{fig2}(c) shows strong exponential tail in the PDF($|\delta r|$) at the lowest Reynolds number. This tail decreases at higher $Re$, until eventually the PDF becomes close to the Gaussian distribution in the highest $Re$ FWT regime.

On a technical note, in the computation of the moments of $|\delta r|$ care should be taken to track fluid particles for sufficiently long time to ensure the convergence of $\gamma_p$. Figure~\ref{fig2}(d) shows the convergence of $\gamma_2$ where the minimal trajectory length is expressed as the tracking time $\tau$ normalized by the Lagrangian integral time $T_L$. The value of $\gamma_2$ converges to 1 for the particles tracked longer than 10$T_L$. Similar convergence tests are also performed for the higher order moments and at different experimental conditions to ensure the converged $\gamma_p$ is obtained.

\begin{figure}
  \centerline{\includegraphics[width=8.5cm]{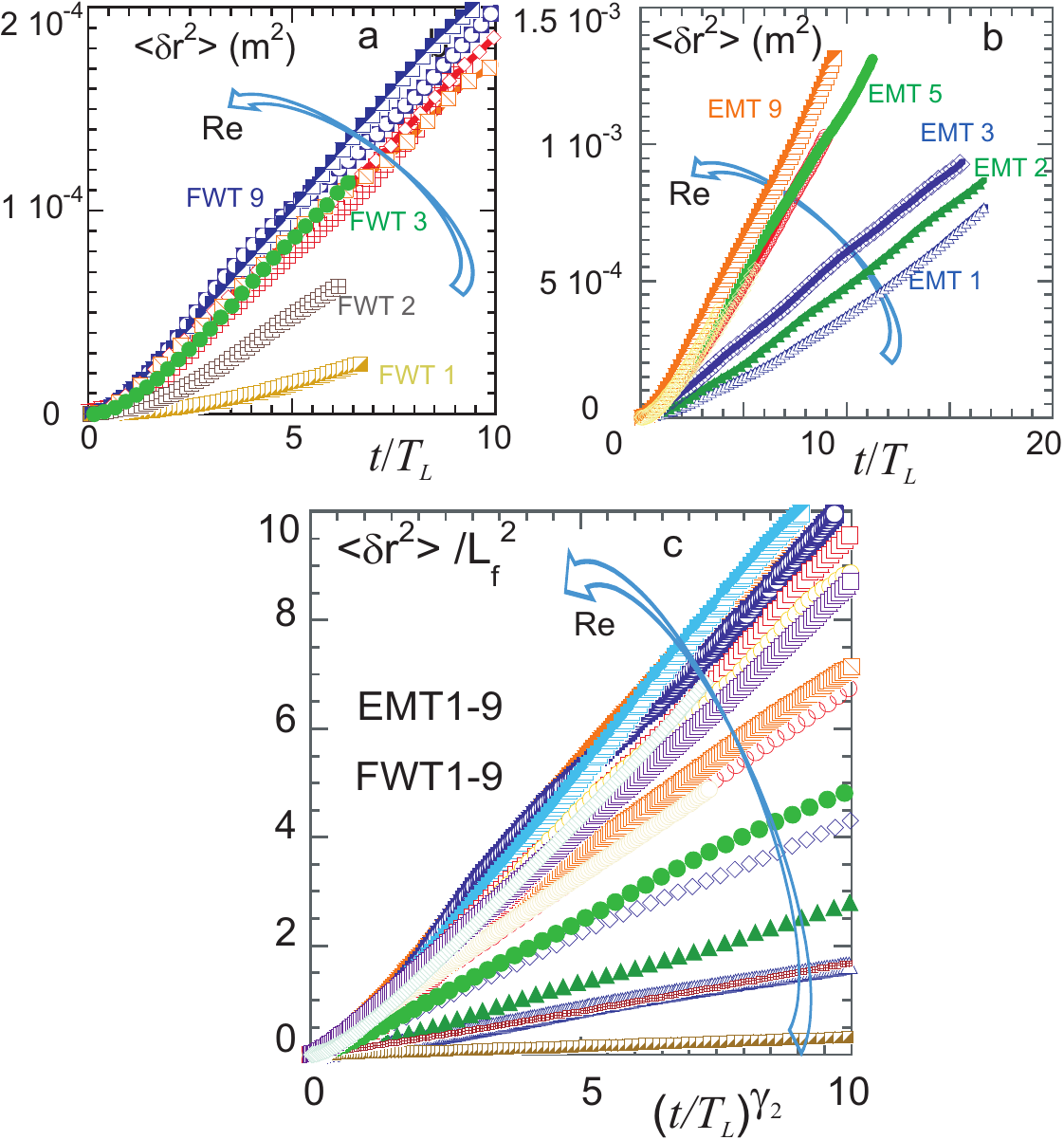}}
  \caption{(a) Mean squared displacement as a function of normalized time $t/T_L$ in FWT1-9. (b) MSD as a function of normalized time $t/T_L$ in EMT1-9. (c) Normalized MSD $\langle (\delta r)^2 \rangle /L_f^2$ versus $(t/T_L)^{\gamma_2}$ in all experiments discussed here. The slopes of the curves are steeper for higher Reynolds numbers, as indicated by the arrows.}
\label{fig3}
\end{figure}

It has been shown in \cite{XiaNC2013} that the MSD $\langle \delta r^2\rangle (t)$ curves measured in turbulence at different forcing levels collapse onto a single line if they are plotted versus the normalized time, $t/T_L$. This is illustrated in Fig.~\ref{fig3}(a) for six FWT regimes at $Re=(56-188)$. At lower $Re$, this is however not the case and the particles are less dispersed, as seen from FWT1-2 in Fig.~\ref{fig3}(a) and in the EMT, Fig.~\ref{fig3}(b).  Also one can see that for the lowest $Re$, the $\langle \delta r^2\rangle (t)$ curves are not precisely linear functions of $t/T_L$. Fig.~\ref{fig3}(c) shows MSD normalized by the square of the forcing scale, $\langle \delta r^2\rangle /L_f^2$, versus normalized time to the power of $\gamma_2$. All curves are reasonably linear, their slopes gradually increase with the increase in the Reynolds number, and they show signs of convergence at high $Re$. If we denote these slopes as $\beta$, we arrive at the expression of Eq.~(\ref{MSD}), namely
$\langle (\delta r)^2 \rangle /L_f^2=\beta(t/T_L)^{\gamma_2}$.

\begin{figure}
  \centerline{\includegraphics[width=8.5cm]{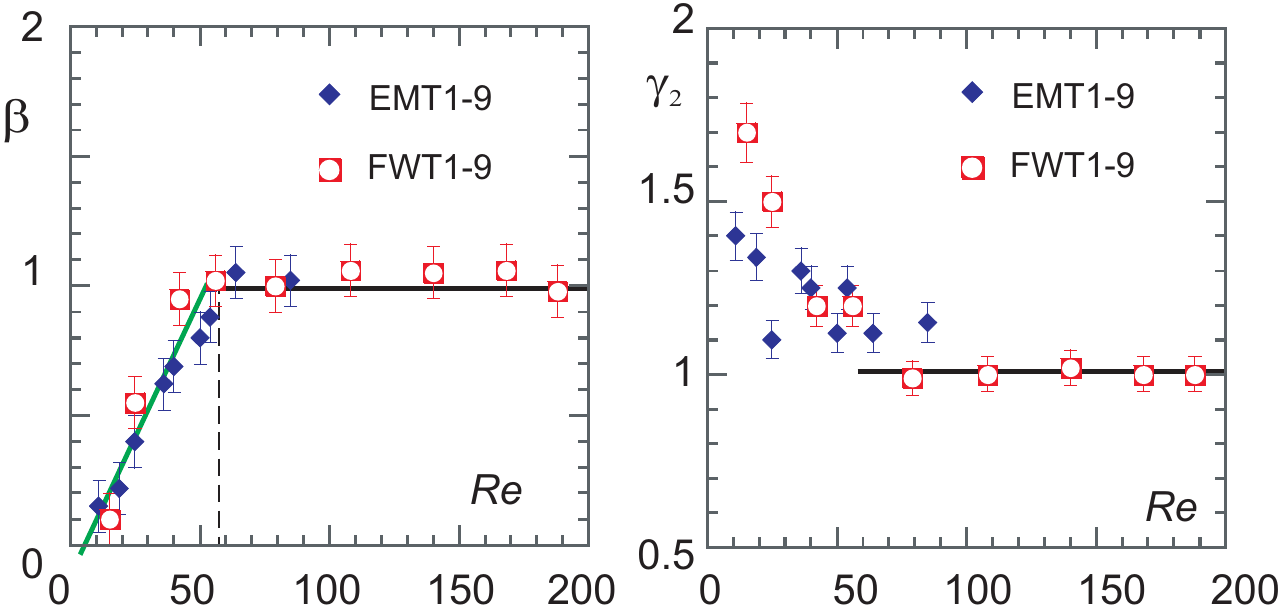}}
  \caption{(a) The anomaly coefficient $\beta$, and (b) the second-order moment of the particle displacements $\gamma_2$ as a function of the Reynolds number for EMT (solid diamonds) and FWT (open squares).}
\label{fig4}
\end{figure}

Figure~\ref{fig4}(a) shows $\beta$, which we refer to as the anomalous dispersion coefficient, versus the Reynolds number for all experiments analyzed here.  Data points from EMT and FWT experiments overlap showing a linear increase in $\beta$ with the increase in $Re$ at $Re<60$. At some critical Reynolds number of $\sim 60$, the anomaly coefficient stabilizes at $\beta=1$ marking the transition to the Brownian-type diffusion. Similarly, $\gamma_2$ converges towards the Brownian diffusion limit of $\gamma_2=1$ with the increase in $Re$, as seen in Fig.~\ref{fig4}(b).

The above results suggest that the anomaly coefficient $\beta$ is a sensitive  parameter which characterizes deviations of single particle dispersion from the Taylor's expression. In the presented experiments $\beta$ varies from 0.1 to 1 and the second-order moment of the particle displacements $\gamma_2$ deviates from the Brownian diffusion value of $\gamma_2=1$ by up to about 60-70\%, at low forcing. $\beta$ appears to be an experimentally accessible, sensitive and reliable new measure of 2D turbulence development. Eq.~(\ref{MSD}) can be viewed as the extension of the Taylor dispersion Eq.~(\ref{Taylor}) into the underdeveloped turbulence and it can be used to quantitatively characterize single particle dispersion from superdiffusive to Brownian motion.

Though it is tempting to conclude from the above results that the Reynolds number plays a crucial role in the turbulence development, and that the kink in Fig.~\ref{fig4}(a) points to the existence of some critical Reynolds number, one should keep in mind the following. $Re$ is defined here using the r.m.s. flow velocity, which is proportional to the kinetic energy stored in the turbulence spectrum ($\langle u^2 \rangle = \int_{k_d}^{k_f} E(k)dk$, where $k_d$ is the large dissipation scale), and the forcing scale $L_f$ which defines the only externally imposed scale of the flow, the scale of the forcing vortices. As has been shown in Fig.~\ref{fig1}, turbulence development coincides with the topological changes in the fluid particle trajectories, namely, with the gradual dominance of flights over traps and the disappearance of the tails in the Lagrangian velocity autocorrelation functions. This leads to the corresponding changes in the higher order moments of the particle displacements, Fig.~\ref{fig2}(a-b), and to the "Gaussianization'' of the PDF of particle displacements, Fig.~\ref{fig2}(c).
As this happens, the ratio of the Lagrangian spatial scale $L_L$ over its Eulerian counterpart $L_f$, converges to a steady value in fully developed turbulence \cite{XiaNC2013}. The degree of turbulence development in 2D flows can be judged from the value of $0 < \beta \leq 1$, with $\beta=1$ marking the state of fully developed 2D turbulence.

\begin{acknowledgments}
This work was supported by the Australian Research Council's Discovery Projects funding scheme (DP110101525). HX would like to acknowledge the support by the Australian Research Council's Discovery Early Career Research Award (DE120100364). HX would like to thank C. Meneveau for useful discussions.
\end{acknowledgments}

\end{document}